\documentclass[nonacm, sigconf]{acmart}

\usepackage{multirow}
\usepackage{tikz}
\usepackage{algorithm}
\usepackage[noend]{algpseudocode}
\usepackage{verbatim}
\usepackage{mathtools}
\usepackage{bm}
\usepackage{wrapfig}
\usepackage{amsmath, amsfonts, amsthm}
\usepackage{textcomp}
\usepackage{booktabs, tabularx}
\usepackage{tikzsymbols}
\usepackage{csvsimple}
\usepackage{xcolor}
\usepackage{soul}
\usepackage{xspace}
\usepackage{listings}
\usepackage{graphicx}
\usepackage{enumitem}
\usepackage{hyperref}
\usepackage[acronym, nohypertypes={acronym}]{glossaries}
\usepackage{tcolorbox}
\usepackage{float}
\usepackage{subcaption}
\newcommand{\uall}{\textbf{uall}\xspace}
\newcommand{\ba}{\textbf{ba}\xspace}
\newcommand{\uten}{\textbf{u10}\xspace}
\newcommand{\eone}{\textbf{e1}\xspace}
\newcommand{\eten}{\textbf{e10}\xspace}
\newcommand{\dbr}{\textbf{d\_br}\xspace}
\newcommand{\rbr}{\textbf{r\_br}\xspace}

\newcommand{\var}[1]{\widetilde{#1}}
\newcommand*{\argmax}{\operatornamewithlimits{argmax}\limits}
\newcommand{\R}{\mathbb{R}}

\renewcommand{\vec}[1]{\boldsymbol{#1}}

\theoremstyle{definition}
\newtheorem{definition}{Definition}

\definecolor{darkred}{rgb}{0.7, 0.0, 0.0}
\definecolor{darkgreen}{rgb}{0.0, 0.7, 0.0}
\definecolor{armygreen}{rgb}{0.29, 0.33, 0.13}

\newcommand{\equationref}[1]{Equation~(\ref{#1})\xspace}

\newcommand{\figref}[1]{Figure~\ref{#1}\xspace}
\newcommand{\secref}[1]{Section~\ref{#1}\xspace}

\newcommand{\tabref}[1]{Table~\ref{#1}\xspace}
\newcommand{\algoref}[1]{Algorithm~\ref{#1}\xspace}

\newcommand{\code}[1]{\texttt{#1}\xspace}
\definecolor{sh_comment}{rgb}{0.12, 0.38, 0.18}
\definecolor{sh_keyword}{rgb}{0.37, 0.08, 0.25}  
\definecolor{sh_string}{rgb}{0.06, 0.10, 0.98} 
\definecolor{KWColor}{rgb}{0.5,0,0.67}
\definecolor{CommentColor}{rgb}{0.15,0.5,0.15}
\definecolor{lightgrey}{rgb}{0.8,0.8,0.8}
\lstdefinelanguage{scala}[]{Java}{
   morekeywords={trait,def,object,with,override,val,type,var} 
}

\lstdefinestyle{Eclipse}{
  xleftmargin=0pt,
  language = scala,
  basicstyle=\small\ttfamily,
  stringstyle=\color{sh_string},
  keywordstyle = \color{sh_keyword}\bfseries,  
  lineskip=-0.0em,
  commentstyle=\color{sh_comment}\itshape,  
  escapeinside={/*@}{@*/},
  numbersep=5pt,
  captionpos=b,
  xleftmargin=0.4cm, xrightmargin=0.5cm,
  morekeywords={invokestatic,invokeinterface,invokevirtual,invokespecial},
}

\makeatletter
\newcommand*{\encircle}[1]{\relax\ifmmode\mathpalette\@encircle@math{#1}\else\@encircle{#1}\fi}
\newcommand*{\@encircle@math}[2]{\@encircle{$\m@th#1#2$}}
\newcommand*{\@encircle}[1]{\tikz[baseline,anchor=base]{\node[draw,circle,outer sep=0pt,inner sep=.2ex] {\textbf{#1}};}}
\makeatother

\acmYear{2021}

\date{}
\begin{document}

\title{Game-Theoretic Malware Detection}
\author{Revan MacQueen}
\affiliation{%
 \institution{University of Alberta}
 \country{Canada}
}
\author{Natalie Bombardieri}
\affiliation{%
 \institution{University of Alberta}
 \country{Canada}
}
\author{James R. Wright}
\affiliation{%
 \institution{University of Alberta}
 \country{Canada}
}
\author{Karim Ali}
\affiliation{%
 \institution{University of Alberta}
 \country{Canada}
}
\begin{abstract}
Malware attacks are costly. To mitigate against such attacks, organizations deploy malware detection tools that help them detect and eventually resolve those threats. While running only the best available tool does not provide enough coverage of the potential attacks, running all available tools is prohibitively expensive in terms of financial cost and computing resources. Therefore, an organization typically runs a set of tools that maximizes their coverage given a limited budget. However, how should an organization choose that set?  Attackers are strategic, and will change their behavior to preferentially exploit the gaps left by a deterministic choice of tools.  To avoid leaving such easily-exploited gaps, the defender must choose a random set.

In this paper, we present an approach to compute an optimal randomization over size-bounded sets of available security analysis tools by modeling the relationship between attackers and security analysts as a leader-follower Stackelberg security game.  We estimate the parameters of our model by combining the information from the VirusTotal dataset with the more detailed reports from the National Vulnerability Database.  In an empirical comparison, our approach outperforms a set of natural baselines under a wide range of assumptions.


\end{abstract}

\maketitle
\section{Introduction}
A recent study by the Ponemon Institute~\cite{ibmsec} shows that malware attacks have been steadily growing with an average cost of US\$4.5 million per attack. To mitigate against such attacks, companies typically run malware detection tools that help security analysts find and eventually resolve potential risks against their running systems. To achieve that, a company may follow one of three approaches. First, a company may run the best tool that they could acquire or develop. However, as the VirusTotal dataset~\citep{VT} shows, no single tool is sufficient to detect all attacks. In the dataset, none of the 86~tools detect more than 84\% of the vulnerabilities in the analyzed malware files. Second, a company may run all possible available tools.  However, this is likely to be impractical due to prohibitive financial costs and limited computing resources. In fact, in their 2019 study, Accenture estimated that only 32\% of companies worldwide deploy discovery and investigation tools due to their rising costs~\cite{accenture}. Third, to achieve the best results given a limited budget, the company may alternatively run a set of tools that can detect most of the potential attack scenarios against the company's systems. But which set of tools should the company choose to run?

A tempting solution is to run the set of tools with the \emph{best historical} coverage (e.g., that would have detected most historical attacks).
However, security is a strategic domain: attackers will change their behavior in response to the defender's actions. Any deterministic choice of tools that does not provide exhaustive coverage will leave gaps that an attacker may exploit. To avoid leaving an easily exploitable attack surface without exhaustively running all available tools, the defender must randomly select the set of tools to run. However, not all randomizations are equal.

In this paper, we present a method for finding the \emph{optimal} randomization over size-bounded sets of tools, under the assumption that the attackers will optimally respond to whichever randomization that the defender chooses. To achieve that, we present a specialized version of a framework of game theoretic models known as Stackelberg security games~\citep{jain}. Prior work has applied the Stackelberg framework to a wide variety of security problems such as airport security~\citep{jain}, preventing fare evasion~\cite{yin}, and detecting and preventing illegal poaching~\citep{yang2014adaptive}.  Our work applies the framework to malware detection and provides an algorithm for finding an optimal randomization over detection tools for the defender.

Modelling any scenario game-theoretically requires a description of the participants' utilities; this is often specified manually by domain experts~\citep[e.g.,][]{jain}.
To estimate the utilities in our model, we use the data from the VirusTotal dataset of malicious files~\citep{VT} in conjunction with information on the impact and difficulty of exploitation of the associated vulnerabilities from the National Vulnerability Database~\citep{NVD}.
This estimate is not intended to be used ``out of the box''; rather, we use it to demonstrate the construction of a utility model from real-world data, and to construct a realistic scenario for evaluating the performance of our proposed model in comparison to several natural benchmarks.  When applying our framework, the utility model can and should be customized to account for individual factors such as the available set of tools, the effectiveness of the tools at detecting different attacks, the severity of undetected attacks, and the relative costs of running tools.

Our empirical evaluation shows that our proposed randomization scheme outperforms a number of deterministic strategies for choosing the \emph{best historical} set of tools, as well as naive uniform randomization schemes.
Some parameters of our utility model represent subjective tradeoffs between different costs such as the relative importance of detecting too many attacks (i.e., false positives) compared to too few (i.e., false negatives).  Our evaluation includes a sensitivity analysis for this subset of parameters that cannot be estimated from data; we find that our results do not qualitatively differ across a wide range of reasonable settings.

To summarize, we make the following contributions:
\begin{enumerate}
    \item Formalize the malware detection problem within the Stackelberg security game framework, and show how to use the mixed integer linear program formulation of~\citet{jain} to find an optimal randomization over sets of tools.
    \item Estimate a realistic utility model for the attacker and defender using data from the National Vulnerability Database~\citep{NVD}, the VirusTotal malware dataset~\citep{VT}, and prior work on false positive rates of malware detection tools~\cite{Measuring}.
    \item Evaluate the performance of our proposed solution using the estimated utility model in comparison to a set of natural baselines.
\end{enumerate}

\section{Background}
\label{sec:background}

In this section, we begin by introducing Stackelberg games.  We then define the solution concept that we use, the strong Stackelberg equilibrium.  Finally, we describe a standard specialization of Stackelberg games for use in the security domain.

\subsection{Stackelberg Games}

Stackelberg games \cite{von2010market} are non-cooperative games between two players: a leader $L$ and a follower $F$.
In a Stackelberg game, the leader chooses a distribution over actions, then the follower chooses their own action distribution, after having observed the leader's action distribution.  An action for each player is then sampled from their chosen distributions, and each player realizes a utility (i.e., payoff) based on the pair of sampled actions.

Unlike normal-form games such as rock-paper-scissors or Prisoner's Dilemma, in which players choose their actions simultaneously, Stackelberg games are played sequentially, with the leader making decisions first and the follower second.  Stackelberg games also differ from extensive-form games such as chess or poker by the timing of the sampling of actions.  In an extensive-form game, one player's action is sampled before the next player begins their turn; thus, the second player's strategy may only condition upon the first player's realized action. In Stakelberg games, the second player may condition on the distribution of the first player's actions.

Each player $i$ has a set $A_i$ of available \emph{actions}.
An \emph{action profile} $(a_L, a_F) \in A_L \times A_F$ is a joint assignment to actions for each player.
A \emph{strategy} $s_i \in S_i = \Delta(A_i)$ for an agent $i$ is a probability distribution over its action set $A_i$, where each action $a_i \in A_i$ is played with probability $s_i(a_i)$.\footnote{We denote the set of all distributions over $X$ with the notation $\Delta(X)$.}
A \emph{pure strategy} is a strategy that assigns probability 1 to a single action, whereas a \emph{mixed strategy} is a strategy that assigns positive probability to multiple actions.
A \emph{strategy profile} ($s_L, s_F) \in S_L \times S_F$ is a joint assignment of mixed strategies for each player.

The outcome of the game is fully determined by the realized action profile.
Each player $i$ has a \emph{utility function} $u_i : A_L \times A_F \to \R$ that maps each action profile to a \emph{utility} that measures how much the agent prefers the outcome, with each agent preferring higher-utility outcomes.
Each player aims to maximize their expected utility, where the expectations are taken over the action distributions.
Players seek only to maximize their own utility function; the utilities of other players are important to a player only insofar as they help to predict those players' actions.

We overload the notation of the utility function to represent expected utility when its arguments are strategies:
\begin{equation}
u_i(s_L,s_F) \doteq \sum_{(a_L,a_F) \in A_L \times A_F} s_L(a_L)s_F(a_F)u_i(a_L,a_F). \label{eq:exp_util}
\end{equation}
Translating our informal description of Stackelberg games to use the notation that we have just defined, a Stackelberg game proceeds as follows:
\begin{enumerate}
    \item The leader chooses a strategy $s_L \in S_L$.
    \item The follower observes $s_L$.
    \item The follower chooses a strategy $s_F \in S_F$.
    \item An action profile $(a_L,a_F)$ is sampled with $a_L \sim S_L$ and $a_F \sim S_F$.
    \item Each agent $i$ receives utility $u_i(a_L,a_F)$.
\end{enumerate}

\subsection{Stackelberg Equilibrium}
\label{sec:stackelberg-eqm}

The utility that a player receives depends upon both its own actions, and the actions of the other player.
We say that an action is a \emph{best response} to the strategy of the other player when its expected utility is maximal given the strategy of the other player.  Formally, we define the set of $i$'s best responses to strategy $s_{-i}$ as
\begin{equation} \label{eqn:br}
    BR_i(s_{-i}) \doteq \{ a_i \in A_i \mid u_i(a_i, s_{-i}) \ge u_i(a'_i, s_{-i}) \quad\forall a'_i \in A_i\}
\end{equation}
where $s_{-i}$ is the strategy of the other player.

In a Stackelberg game, a profile in which both players are behaving optimally is called a Stackelberg equilibrium.
There can be multiple such equilibria in a game, especially when the follower is indifferent between multiple actions given the strategy of the leader.  In this paper, we focus on a particular refinement of Stackelberg equilibrium called the strong Stackelberg equilibrium (SSE), in which the attacker breaks ties in favor of the defender~\cite{von2004leadership}.

\begin{definition}[Strong Stackelberg Equilibrium]
    A strategy profile $(s^*_L, s^*_F)$ is a \emph{strong Stackelberg equilibrium} if it satisfies the following:
    \begin{enumerate}
        \item $u_L(s^*_L, BR_F(s^*_L)) \ge u_L(s'_L, BR_F(s'_L)) \quad\forall s'_L \in S_L$,
        \item $s_F(a_F) > 0 \implies a_F \in BR_F(s^*_L) \quad\forall a_F \in A_F$, and
        \item $u_L(s^*_L, s^*_F) \ge (s^*_L, a_F) \quad\forall a_F \in BR_F(s^*_L)$.
    \end{enumerate}
\end{definition}

It may seem implausible to assume that the follower would break ties in favor of the leader; after all, does the follower not only care about their own utility?  Would it not be more conservative to assume that the follower will break ties by choosing the best response with the \emph{lowest} utility to the leader?\footnote{In fact, this assumption is sometimes made.  A strategy profile in which the follower breaks ties to minimize the leader's utility is called a \emph{weak} Stackelberg equilibrium.}
In fact, by changing their action distribution by an arbitrarily small amount, the leader can always cause the follower to strictly prefer one of their best responses. Since this change is very small, the utility cost to the leader is also infinitesimal.  Thus, assuming that the follower will choose the best-response which is optimal for the leader is essentially without loss of generality.

\subsection{Security Games }
\label{sec:SecurityGames}

Security games are a frequently used framework in which security situations are modeled as a Stackelberg game \cite{Avenhaus}.  The two players are the \emph{Defender}, who plays the leader role, and the \emph{Attacker}, who plays the follower role.
%
In a security scenario, the attacker can typically observe the defender's actions (e.g., randomized patrols) over time before choosing their own attack, allowing the attacker to estimate the defender's random distribution over actions.  This setup motivates the use of Stackelberg games for modeling these scenarios.

In this paper, we use the \emph{compact security games} framework of \citet{securitygames}.
The defender seeks to defend a set of \emph{targets} $T=\{t_1, \ldots, t_n\}$ using a set of \emph{resources} $R=\{r_1,\ldots,r_m\}$.  The attacker's action set consists of the targets; they choose a target or distribution over targets to attack.  The defender's actions set consists of the resources; they choose a resource or distribution over resources before the attacker chooses an action.

A ``resource'' in the model may represent multiple actual real-world resources.
A single resource is intended to model a plan that the defender could feasibly execute.
For example, in the airport security domain, \citet{jain} represent a single feasible schedule of patrols and placements of vehicle checkpoints as a resource.  As we describe in Section~\ref{sec:sg-for-malware}, we represent a set of malware detection tools to run as a single resource.

Each of the Defender's resources $r$ induces a \emph{coverage vector} $\vec{p^r} \in [0,1]^n$, with each element $p^r_t$ representing the probability that target $t$ will be ``covered''.  When the attacker attacks a covered target, the attack fails; otherwise, the attack succeeds.  When an attack on target $t$ succeeds, the attacker receives utility $u^1_a(t)$ and the defender receives utility $u^1_d(t)$; when an attack on target $t$ fails, the attacker receives utility $u^0_a(t)$ and the defender receives utility $u^0_d(t)$.  These utilities are not constrained to be zero-sum; i.e., it is not necessarily the case that $u^1_d(t) = -u^1_a(t)$.

The utility of a player $i$ when the attacker attacks target $t$ and the defender chooses a resource $r$ with coverage vector $\vec{p}^r$ is thus

\begin{equation}
    u_i(\vec{p}^r,t) = p^r_tu_i^0(t) + (1-p^r_t)u_i^1(t).
\end{equation}
When the defender chooses a mixed strategy $s_d \in \Delta(R)$ and the attacker chooses a mixed strategy $s_a \in \Delta(T)$, the expected utilities are computed in the straightforward way as
\begin{equation}
 u_i(s_d,s_a) = \sum_{r \in R}\sum_{t\in T} s_d(r)s_a(t)u_i(\vec{p}^r, t).
\end{equation}
\section{Security Games for Malware Detection}
\label{sec:sg-for-malware}

Due to both time and resource constraints, it is not practical to use every available detection tool to defend against malware.
Nonetheless, if the defender always uses the same subset of the available tools, then malware authors may target vulnerabilities that the particular set of tools does a poor job of detecting.  We assume that the attacker is \emph{strategic}, in the sense that they will adjust their behavior in response to the actions of the defender. For this reason, simply using a tool (or set of tools) that performs well on the historic distribution of attacks is not sufficient.

Our work applies the security games framework to malware detection.
Although our model assumes that there may be many non-colluding attackers, the model represents attacks from different attackers as separate games, each with a single attacker. This is why we refer to each game having a single attacker, even though there would likely be more than one attacker in reality. However, we assume all attackers have the same utility function and, therefore, the optimal strategy for the defender generalizes over all games.

In our model, the organization/company plays the role of the defender in the security game, while a potential hacker is the attacker.  The attacker seeks to upload a file that exploits a security vulnerability; the defender aims to prevent this exploitation, while still providing access to their services. The attacker's action set consists of a set $V$ of vulnerabilities to target.

The defender has a set of malware detection tools available, each with associated probabilities of detecting attacks on different vulnerabilities (as well as false positive rates).  We constrain the maximum number of tools that the defender can use in each schedule by a budget $b$, the number of tools that they are able or willing to run. The set $S$ of resources thus consists of all subsets of detection tools containing $b$ or fewer tools.  We refer to these subsets as \emph{schedules} of detection tools, with a \emph{budget} of $b$.  The defender chooses a distribution over schedules before the attacker chooses which vulnerability to attack. The goal with this approach is to determine an optimal distribution, which is chosen once, and then sampled from without further changing it.

\subsection{Utility Model}
\label{sec:utility}

Each schedule $s$ has an associated probability $p(s,v)$ of detecting an attack targeting $v$, and an overall cost $c_d(s)$ of running the schedule.  In our empirical evaluation, we estimate the detection probabilities and false positive rate of a schedule based on the detection probabilities and false positive rates of the underlying tools, and construct a cost based on the false positive rate. Other costs such as resource requirements and runtimes can easily be included when implementing a model such as the one we present, but such data was not available to us.

Different vulnerabilities have different impacts on the defender when attacked without being detected.  We say that a defender receives a negative ``reward'' $r_d(v)$ for failing to detect an attack on vulnerability $v$.  The attacker receives a reward $r_a(v)$ for successfully attacking $v$, and incurs a cost $c_a(v)$ for attacking $v$.
We estimate the cost $c_a$ based on the exploitability of the vulnerability (defined in Subsection~\ref{subsec:impact_difficulty})  . We estimate the reward $r_a(v)$ to the attacker and negative reward $r_d(v)$ to the defender of a successful attack on $v$ based on the impact of the exploit.

The attacker's utility from attacking vulnerability $v$ when the defender chooses schedule $s$ is
\begin{equation}
    u_a(s,v) \doteq (1-p(s,v))r_a(v) - \gamma_a c_a(v) ,\label{eq:attacker_util}
\end{equation}
where $\gamma_a$ represents the relative weighting of the units of reward and the units of cost.
The attacker incurs the cost $c_a(v)$ regardless of whether the attack is detected, but only receives the reward $r_a(v)$ if the attack is not detected.

The utility for the defender is
\begin{equation}
    u_d(s,v) \doteq (1-p(s,v))r_d(v) - \gamma_d c_d(s), \label{eq:defender_util}
\end{equation}
where $\gamma_d$ again trades off units of cost and units of reward. We assume the reward for a detected attack to be zero for the defender.
Equations~\eqref{eq:attacker_util} and \eqref{eq:defender_util} differ from \equationref{eq:exp_util} because
they specify the utility of an action profile (i.e., a schedule chosen by the defender, and a vulnerability chosen by the attacker),
whereas \equationref{eq:exp_util} takes expectation over these action profile utilities to compute the expected utility of a profile of mixed strategies.

\subsection{Finding Optimal Randomized Strategies} \label{subsec:find_optimal_strats}
To solve for a strong Stackelberg equilbrium (SSE) for the game defined above, we encode the game as a mixed integer linear program (MILP)~\citep{jain}.
By solving the MILP, we efficiently compute the optimal schedule randomization for the defender, assuming that the attacker will best respond. We first specify the MILP in Equations~\eqref{eq:maximization}--\eqref{eq:attacker-br}, translated from the notation of \citet{jain} to our own. We then describe the meaning of each constraint.


\begin{align}
    \text{maximize}  \quad   &\var{U_d}\label{eq:maximization}\\
    \text{subject to}\quad
    & \var{y}_v \in \{0, 1\}    \quad \forall v \in V \label{eq:deterministic-attack}\\
    & \sum_{v \in V} \var{y}_v = 1 \label{eq:y-valid-probability}\\
    & 0 \le \var{p}_s \le 1         \quad \forall s \in S \label{eq:x-valid-probability1}\\
    & \sum_{s\in S} \var{p}_s = 1 \label{eq:x-valid-probability2} \\
    & \var{U_d}       - \sum_{s \in S}\var{p}_su_d(s,v) \le (1-\var{y}_v)Z \quad \forall v \in V  \label{eq:defender-br}\\
    0 \le &\var{U_a} - \sum_{s \in S}\var{p}_su_a(s,v) \le (1-\var{y}_v)Z \quad \forall v \in V \label{eq:attacker-br}
\end{align}

The defender's decision variables are the set $\var{p}_s \;\forall s \in S$,
where $\var{p}_s$ gives the probability of running schedule $s$.  Constraints \eqref{eq:x-valid-probability1} and \eqref{eq:x-valid-probability2} ensure that the vector $\var{p}$ is a valid probability distribution.

The attacker's decision variables are the set $\var{y}_v \;\forall v \in V$,
where $\var{y}_v$ gives the probability of attacking vulnerability $v$.  Constraints \eqref{eq:deterministic-attack} and \eqref{eq:y-valid-probability} ensure that exactly one element of $\var{y}$ will be set to 1, with all others set to 0; i.e., the Attacker is constrained to deterministically attack a single vulnerability.  As discussed in \secref{sec:stackelberg-eqm}, this is essentially without loss of generality.

Constraint~\eqref{eq:attacker-br} forces the attacker strategy $\var{y}$ to be a best response to the defender's strategy, as follows.  First, for the vulnerability $v^*$ that is exploited, $\var{U_a}$ must exactly equal the attacker's expected utility $\sum_{s \in S}\var{p}_su_a(s,v^*)$, because the difference between $\var{U_a}$ and the attacker's expected utility must be both weakly greater and weakly less than zero.  Second, for every other vulnerability $v$ that is not exploited, the same variable $\var{U_a}$ must be weakly greater than the expected utility of attacking $v$; i.e., there must not be any other vulnerability with a strictly greater expected utility, given the strategy of the defender.  The constant $Z$ is an arbitrary large value; its presence allows us to upper bound the difference between $\var{U_a}$ and expected utility for $v^*$ only, while removing the upper bound for the other vulnerabilities.

Constraint~\eqref{eq:attacker-br} serves a similar role for the defender's expected utility: The variable $\var{U_d}$ is required to be weakly less than the defender's expected utility given its mixed strategy $\var{p}$ and the vulnerability chosen for attack.  Since $\var{U_d}$ is the maximization objective, there is no need for a lower bound to force equality.

\section{Estimating Parameters of Our Utility Model}
\label{sec:parameters}

Our utility model has a number of free parameters that we need to specify: the set $V$ of vulnerabilities, the set $S$ of schedules, the probabilities $p(s,v)$ of detection, the costs $c_a$ and $c_d$, the rewards $r_a$ and $r_d$, and the tradeoffs $\gamma_a$ and $\gamma_d$ between rewards and costs.  For the purposes of evaluating the security game framework empirically, we estimated these parameters using a malware dataset made available by VirusTotal (VT)~\cite{VT}.
The dataset provides over 30,000 live malware samples, as well as results from associated scans with 86 antivirus tools.
Each individual scan contains information such as the tool version, whether it flagged the file as malicious, and a label describing why the file was flagged, the file type, tags, other submissions of the same file, and how many tools flagged the file as malicious.

The ``tags'' section of the VT reports may include one or more CVE tags, which are unique identifiers that differentiate software vulnerabilities. These tags allow us to determine which CVEs are targeted among the files present in VT, allowing us to map out the possible attacks that an attacker may attempt, representing all of the attack targets.  We let the set $V$ of vulnerabilities be the set of all CVEs that are tagged on at least one file in the VT dataset.

The set of 86 antivirus tools used in the VT dataset is the set of underlying tools available to the defender.  We evaluate schedules with size limits varying from $b=1$ to $b=4$ tools.

\subsection{Impact and Difficulty} \label{subsec:impact_difficulty}

CVEs represent different classes of software vulnerabilities, and in our case, different attack targets. We use this information to relate the vulnerabilities to the National Vulnerability Database (NVD)~\cite{NVD}.
For each unique CVE number, NVD provides a detailed report containing more information about the vulnerability. This information includes two parameters for each CVE, which we use in our model: \emph{impact score} and \emph{exploitability score}.

The impact score represents the severity of the exploited vulnerability. NVD calculates this score by numerically quantifying three factors; the degree to which confidential information is made available to the attacker, the degree to which the affected device or network can be modified by a successful attack, and the degree to which access to the affected device or network is interrupted/prevented~\cite{NVDcalc}. The numbers are multiplied together with a constant to give a rating out of 10, with higher numbers representing more dangerous exploits.
For each CVE $v$, we set the reward $r_a(v)$ that the attacker receives in the case of a successful attack to be the impact score of the CVE.  We correspondingly set $r_d(v) = -r_a(v)$. However, our final utility model is not be zero-sum, because the costs of the attacker and the defender differ.

The exploitability score represents the difficulty of carrying out an attack on the vulnerability. This score is calculated by numerically quantifying four factors: how remotely the attack can be made, the conditions and knowledge that must be present for the attack to be successful, the degree of privilege that the attacker must have to be successful, and the degree of interaction required by a user of the targeted system~\cite{NVDcalc}. These numbers are then multiplied together with a constant, which gives a rating out of 10, with higher numbers representing vulnerabilities that are harder to exploit. For each CVE, our model uses this number as the attacker's cost $c_a(v)$ for carrying out an attack on that vulnerability.

\subsection{Detection Probabilities}

To estimate the detection probability of a tool detecting each vulnerability, we count the number of files in the VT dataset tagged with the vulnerability's CVE that the tool flags as malicious. We then divide that count by the number of files tagged with the CVE on which the tool was run.\footnote{Not every tool was run on every file.}
We do not require that the tool report the ``correct'' vulnerability, because each tool typically reports an opaque, tool-specific vulnerability code that often does not directly correspond to a CVE.

Estimating the detection rate for a schedule containing more than a single tool is more complicated, because the detection probabilities for multiple tools are, in general, not independent. This observation is consistent with the findings of \citet{Measuring} who did a similar experiment using clustering. The authors found that 16 of the tools in VT are highly related, with one cluster of 6~tools having nearly identical labels for 98\% of the dataset.

We estimate the detection rates of schedules in a similar manner to how we calculate the single tool detection rates.
For a given schedule $s$ and vulnerability $v$, we first find the set of all files tagged with the vulnerability's CVE on which \emph{at least one} of the schedule's tools were run.  We then estimate the detection rate $p(s,v)$ as the number of files in this set that were flagged as malicious by \emph{any} of the schedule's tools, divided by the total size of the set.
This approach accounts empirically for the dependence in tools' detection rates, although it potentially underestimates a schedule's detection rate. This is because a file on which not every tool of the schedule was run which was flagged as benign might have been flagged as malicious by one of the omitted tools.

\subsection{False Positives as Defender Costs}

The cost $c_d(s)$ to the defender of running a schedule $s$ may incorporate a number of factors such as runtime, memory requirements, and false positive rates. For the purposes of our empirical evaluation, we assume that all schedules have approximately the same resource requirements. Therefore, we calculate $c_d(s)$ based solely on an estimate of the false positive rate.

False positives can pose a high potential cost. In VT, the majority of tools the dataset returned false positives when tested on a benign set of obfuscated files, with about 25\% of tools tagging over 70\% of benign files scanned as malicious~\cite{Measuring}. Some tools may have extremely high detection rates not because they are good at finding malware, but because they are too ``trigger-happy,'' or are overly sensitive when scanning for viruses (i.e., they have a high false positive rate in addition to a high true positive rate). 

To estimate the false positive rate for every tool, we used the dataset provided by \citet{Measuring}. To achieve that, we divided the number of times that a tool gave a false positive by the number of times that a tool was run on benign files. 
For schedules of sizes greater than one, we divide the number of times that at least one tool in that schedule gave a false positive by the number of files that include at least one tool from the schedule.
This calculation gives us an overall false positive rate for each schedule, which we use as the schedule cost $c_d(s)$.


\subsection{Reward-Cost Tradeoff}
Since we parameterize our model using a variety of scores from NVD and elsewhere, there is no clear method of comparing one unit of reward against one unit of cost. For example, the attacker's costs are given by the  NVD exploitability score and its rewards are given by the impact score. We have no prior knowledge how an attacker may weight the relative importance of each score. Instead, our model contains the parameters  $\gamma_a$ and $\gamma_d$, defined in Equations \eqref{eq:attacker_util} and \eqref{eq:defender_util}. These values represent the trade-off between costs and rewards for the attacker and defender. A relatively low value of $\gamma$ corresponds to a player who values a unit of reward much more than a unit of cost. A high value is for a player who is deeply concerned about costs. These parameters may be tuned to represent different types of players. For example, if $\gamma_a = 0$, then the attacker's utility does not account for costs, analogous to an attacker with much more resources than needed for any attack. On the other hand,  $\gamma_a = 10$  represents a very cost-cautious attacker. We present the effects of changing $\gamma_a$ and $\gamma_d$ in \secref{subsec:results_gamma}.


\subsection{Limitations}
Our approach to estimating the parameters of our model has a few limitations:

\begin{enumerate}
    \item \textbf{Limited CVE Availability:}
          Although the VT dataset includes over 30,000 files that span tens of platforms, only a fraction of these files (3,515) have corresponding CVE tags, limiting us to those files.
    \item \textbf{False Positives:}
          The VT dataset does not include benign files, making it impossible to estimate false positive rates.  We thus estimated false positive rates using a dataset provided by~\citet{Measuring}.
    \item \textbf{Low Occurrence Rates:}
          Some CVE samples occur very infrequently in the VT dataset, which may lead to extreme estimates of detection probabilities close to either 0 or 1.
          We mitigate this effect by adding \emph{pseudocounts} to the underlying data; this leaves the estimates of detection probabilities for CVEs with many occurrences largely unaffected, while normalizing the detection probabilities of CVEs with low occurrence counts to be closer to uniform. We give the probability of a schedule $s$ detecting an attack against vulnerability $v$ as:
       \begin{align}
              p(s \ \text{detects} \ v)  &= \dfrac{\text{ \# files where $s$ detects $v$}  + n }{\text{ \# files with $s$ and $v$} + 2n},
        \end{align}
        where $n$ is a parameter determining the weight given by the prior to a uniform distribution. We use $n=2$.
          \end{enumerate}

\section{Evaluation}
\label{sec:evaluation}

We evaluate our model through the following research questions:
\begin{itemize}
    \item[] \textbf{RQ1:} How does our proposed method of randomization compare to the baseline strategies?
    \item[] \textbf{RQ2:} Does changing the reward-cost tradeoff for the attacker ($\gamma_a$) or the defender ($\gamma_d$) change our results?
\end{itemize}







\subsection{Experimental Setup}
We conducted all our experiments on a system with 2 AMD EPYC 7351 2.4~GHz 16-core processors with 32~KB L1 data cache, 64~KB L1 instruction cache, 512~KB L2 cache and  8,192~KB L3 cache, 472~GB of RAM and running Ubuntu 18.04.4 LTS. We used IBM~CPLEX~12.10 for solving the MILP.

\algnewcommand{\LineComment}[1]{\State \(\triangleright\) #1}
\begin{algorithm}[t]
\caption{Compute Deterministic Best Response}\label{alg:prune_dominated}
\begin{algorithmic}[1]
    \LineComment $BR_a(s)$:= attackers best response $\forall s \in S$
    \For{$s \in S$}
    \State $BR_a(s) \gets \argmax_{v\in V} u_a(s, v)$   
    \EndFor
    \State $s^*  \gets \argmax_{s\in S} u_d(s, BR_a(s))$  \Comment{defender best response}
    \State $v^* \gets BR_a(s^*)$ \Comment{attacker best response to $s^*$}
    \State \Return $u_d(s^*, v^*), u_a(s^*, v^*) $ \Comment{utilities for players}
\end{algorithmic}
\label{alg:solve_d_br}
\end{algorithm}

To assess the effectiveness of optimally randomized malware detection, we evaluate our method against a number of other strategies that the defender may use:

\begin{enumerate}
    \item \label{strat:uniform} \textbf{Uniform Randomization:} Defender uniformly randomizes over all schedules in $S$.

    \item \label{strat:best_avg} \textbf{Best Average Detection Rate:} The defender always runs the schedule $s'$ with the best average detection rate.
    \begin{align}
        s' \in \argmax_{s\in S} \frac{1}{ |V|} \sum _{v \in V}p(s \ \text{detects} \  v)
    \end{align}

    \item \label{strat:rand_best_avg} \textbf{Randomized Best Average Detection Rate:}  Extension of best average to uniformly randomize over the $m$ strategies with highest average detection rates.

    \item \label{strat:best_util} \textbf{Highest Expected Utility:}
    The defender chooses a schedule $s'$ with the highest expected utility, calculated by taking the expectation of utility over the likelihood of vulnerabilities being exploited.
    \begin{align}
        s' \in \argmax_{s\in S} \sum _{v \in V} u_d(s,v) \cdot p(v), \label{eq:expect_util_strategy}
    \end{align}
    where the probability of each CVE is given by its occurrence frequency in the VirusTotal database.

    \item \label{strat:random_best_util} \textbf{Randomized Highest Expected Utility:}
    As with best average detection rate, we extend the highest expected utility strategy to allow for uniform randomization over the $m$ strategies with the highest values of \equationref{eq:expect_util_strategy}.

    \item  \label{strat:deterministic_br}\textbf{Deterministic Best Response:} The defender best responds to the attacker by only choosing a single schedule $s^*$. Unlike the above strategies, deterministic best response takes into consideration the actions of the attacker. \algoref{alg:solve_d_br} gives the procedure for finding $s^*$, the attacker's best response $v^*$ to $s^*$, and the utilities for each player. This baseline is a restricted case of our proposed strategy (\secref{subsec:find_optimal_strats}), where the defender assigns probability 1 to $s^*$ and 0 to all other schedules. 
\end{enumerate}

\subsection{Comparison to Baselines (RQ1)}

\begin{figure}
\centering

\label{fig:legend}
\includegraphics[width=\columnwidth]{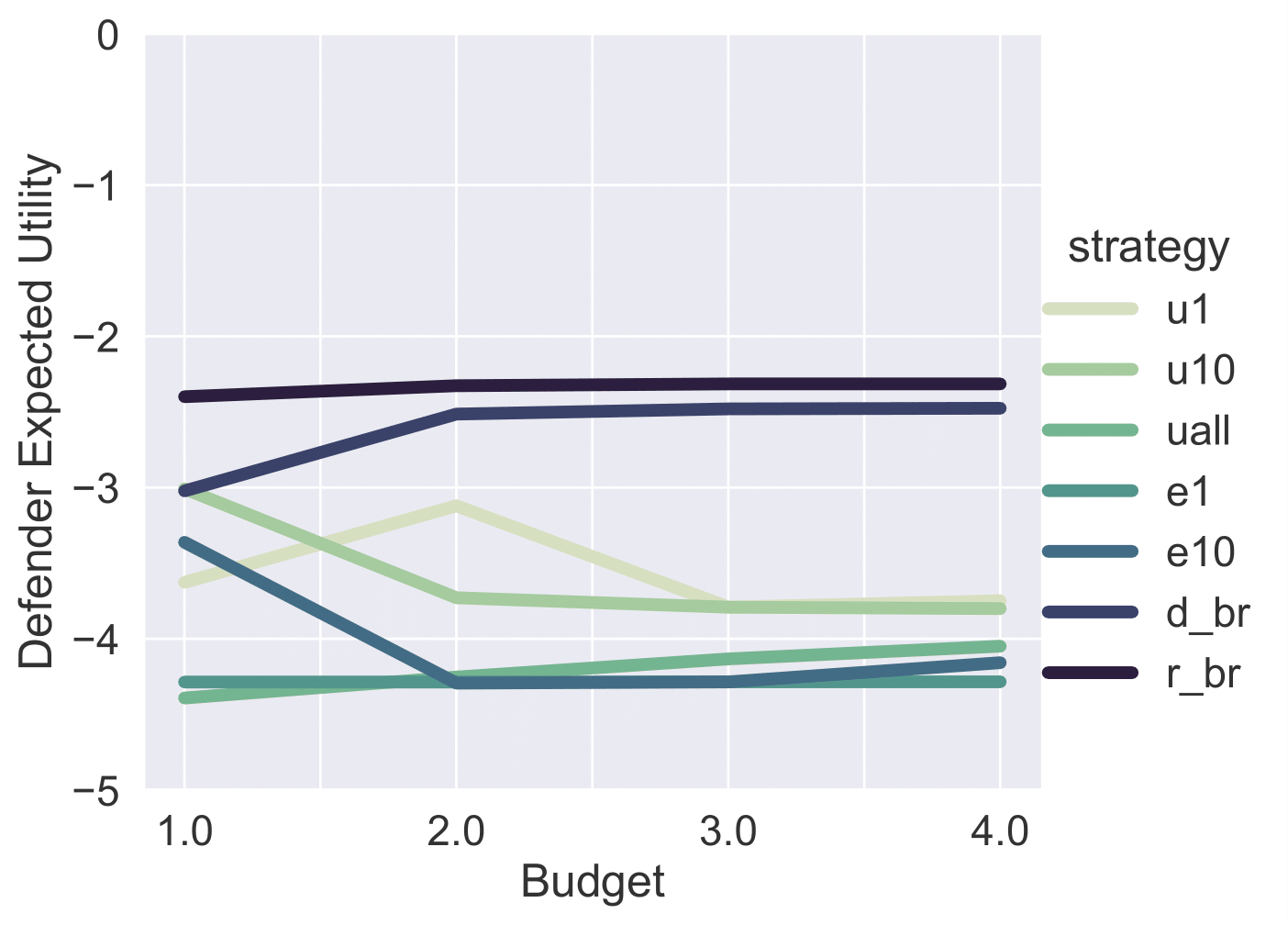}
\caption{Comparing our proposed randomized best response strategy (\rbr) to various baseline strategies. Reward-cost tradeoffs are set to $\gamma_a$ = 1 and $\gamma_d = 2$ for the attacker and the defender, respectively. Utilities closer to~0 indicate better performance. Our proposed method of optimal randomization \rbr outperforms all baselines.}
\label{fig:results}
\end{figure}

We evaluate the following strategies: best average detection rate (\textbf{ba}), randomized best average detection rate where $m$ = 10 (\textbf{u10}),  uniform randomization (\textbf{uall}), highest expected utility (\textbf{e1}), randomized highest expected utility where $m$ = 10 (\textbf{e10}), deterministic best response (\textbf{d\_br}), and our proposed randomized best response (\rbr). In all cases, the defender has a negative expected utility because the defender cannot ``win'' this game, they only seek to minimize their losses. A perfect strategy in a world without costs, detects all attacks with no false positives, and gives the defender a utility of~0. Therefore, the closer to~0 that a strategy's utility is, the better it performs.

\figref{fig:results}\label{labeldescriptions} shows the main results of the comparison. The $x$-axis is the the defender budget $b$. The $y$-axis represents the expected utility of the defender for each strategy. Regardless of the budget, randomized best response (\rbr) outperforms all other strategies. The difference between randomized best response and deterministic best response (\dbr) shows the benefit that the defender realizes through randomization. This is because a deterministic best response  is more easily exploited by the attacker than a randomized best response. For the remaining strategies, the defender does not take into consideration the actions of the attacker and thus generally earns a much lower utility.

For budgets $b \in \{1,2,3, 4\}$, \tabref{table:d_br_schedules} shows the deterministic best response and the associated expected utility, while \tabref{table:r_br_schedules} shows the full randomized best response and expected utility. If the defender chooses a randomized best response, the expected utility for a budget of 3 (-2.269) is only a marginal improvement over the utility for a budget of 2 (-2.270). A budget of 4 yields no additional utility over 3, implying that increasing the budget beyond 2 yields diminishing results for the defender. In other words, given the optimal randomization that our approach computes, the defender achieves the same coverage using fewer tools.

\begin{tcolorbox}
\textbf{RQ1:} Randomizing and considering the actions of attackers increase defender utility. For all schedule sizes, our model (\rbr) outperforms all baselines.
\end{tcolorbox}

\begin{table}
\centering
\caption{Deterministic best response: schedules and probabilities ($\gamma_a = 1$, $\gamma_d = 2$). For each budget, we show the defender utility and the chosen schedule. Since this strategy chooses one schedule, the probability for each chosen schedule is~1.}
\begin{tabularx}{\columnwidth}{crX}
\toprule
\textbf{Budget} & \textbf{Utility} & \textbf{Schedule} \\ \midrule
1 & -3.528 & F-Secure \\ \midrule
2 & -2.601 & TheHacker, ZoneAlarm \\ \midrule
3 & -2.599 & DrWeb, Kaspersky, TheHacker \\ \midrule
4 & -2.591 & Antiy AVL, ESET NOD32, NANO Antivirus, TheHacker\\ \bottomrule
\end{tabularx}
\label{table:d_br_schedules}
\end{table}

\subsection{Does Varying $\gamma_a$ and $\gamma_d$ Affect Results? (RQ2)} \label{subsec:results_gamma}
Next we consider whether changes in the attacker's and defender's cost-reward trade-offs ($\gamma_a$ and $\gamma_d$, respectively) affects our results. By varying $\gamma_d$ in \equationref{eq:attacker_util}, we vary the importance of detection cost to the defender, which in our model is the false positives rate. Varying $\gamma_a$ in \equationref{eq:defender_util} controls the relation between impact and exploitability for the attacker.

Since these parameters cannot be estimated from data, we empirically study the effect of varying them on \rbr. To achieve that, we consider two scenarios: first we hold $\gamma_d$ constant and vary $\gamma_a$, then we hold $\gamma_a$ constant and vary $\gamma_d$.

\textbf{\textbf{}}In the first experiment, we set $\gamma_d$ to be~2 and vary $\gamma_a$ to be a value of $\{0, 0.25, 0.5, 0.75, 1, 2 \}$. We chose non-negative values for $\gamma_a$ because values less than~0 do not make sense in our model, as this would give the attacker a bonus for choosing a more costly attack. We upper-bounded $\gamma_a$ at~2; values higher than~2 leave the attacker with a negative utility, because the costs outweigh the benefits of attacking. An attacker with a negative utility for any given attack would certainly choose not to attack, thus preventing this game from ever taking place.

\begin{figure*}
    \centering
         \centering
     \begin{subfigure}[b]{\textwidth}
         \centering
         \includegraphics[width=\textwidth]{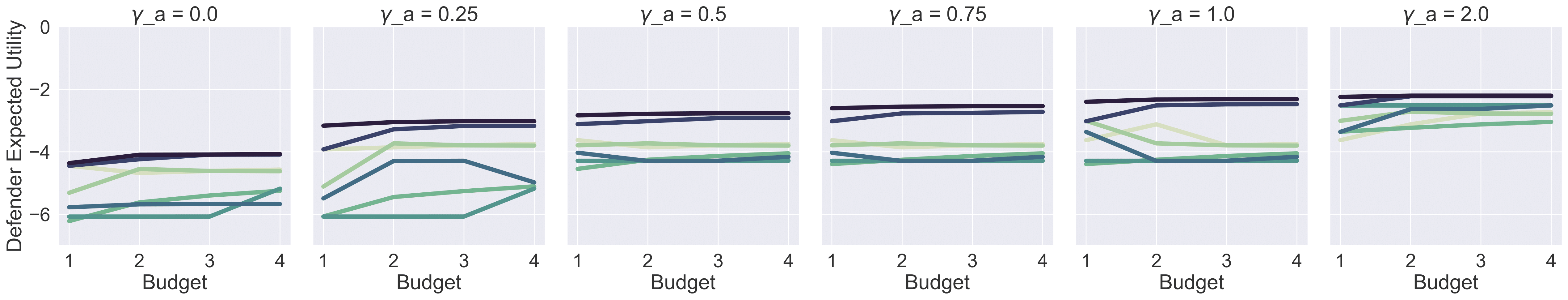}
         \caption{$\gamma_d=2$, varying $\gamma_a$}
     \end{subfigure}
     \hfill
     \begin{subfigure}[b]{\textwidth}
         \centering
         \includegraphics[width=\textwidth]{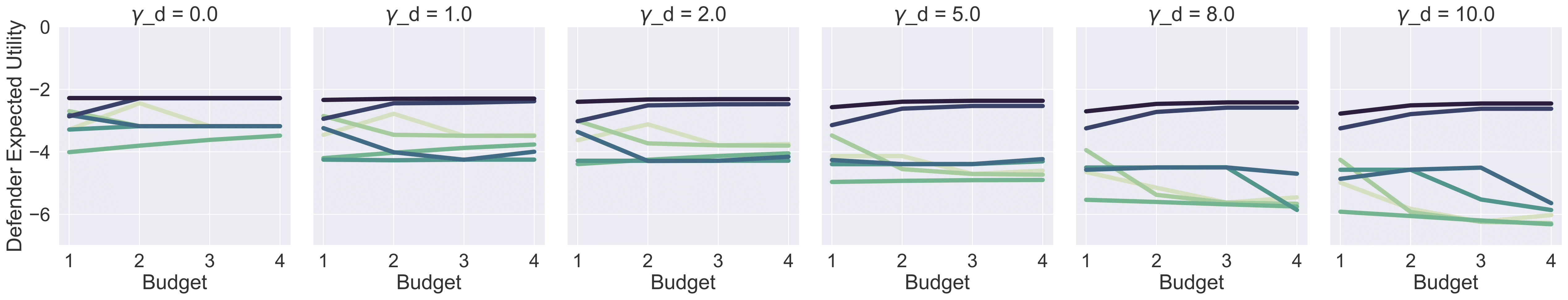}
         \caption{$\gamma_a=1$, varying $\gamma_d$}
     \end{subfigure}
     \hfill
     \begin{subfigure}[b]{0.6\textwidth}
         \centering
         \includegraphics[width=\textwidth]{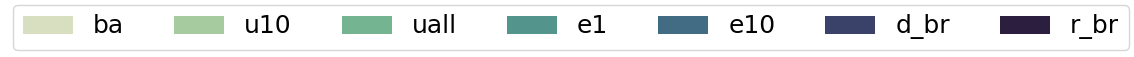}
     \end{subfigure}
    \caption{Expected utility of the defender with varying $\gamma_d$ and varying $\gamma_a$. On the first row, we fix $\gamma_d=2$ and vary $\gamma_a$. On the second row, we fix $\gamma_a=1$ and vary $\gamma_a$. We compare our proposed randomized best response (\rbr) to deterministic best response (\dbr) and other baselines, described in \secref{labeldescriptions}. Utilities closer to zero indicate better performance. In all cases \rbr outperforms all baselines. }
    \label{fig:results_yds_yas}
\end{figure*}

For each setting of $\gamma_a$, we compute the optimal strategy \rbr and all the baselines. The first row of \figref{fig:results_yds_yas} shows the results for the defender's utility in each case. Each subplot shows one setting of $\gamma_a$. Our results show that varying $\gamma_a$ does not change the fact the \rbr is optimal in every case and still outperforms all baselines. 

In the second experiment, we vary $\gamma_d$ using a set of reasonable values: $\{0, 1, 2, 5, 8, 10\}$.
We restrict our attention to non-negative values for $\gamma_d$, because a negative value incentivizes the defender to choose schedules with a high false positive rate. We chose~10 as an upper bound for $\gamma_d$ because that threshold implies that the maximum penalty for failing to detect an attack (a 10 in the NVD impact score) is equal to the maximum cost the defender may incur, which occurs when the false positive rate is $1.0$.
The second row of \figref{fig:results_yds_yas} shows the utility of \rbr and all baselines under different settings of  $\gamma_d$. Similar to our first experiment, changing $\gamma_d$ does not affect the optimality of \rbr, which still outperforms all baselines. Moreover, the results show that increasing $\gamma_d$ greatly diminishes the performance of many of the baselines. This is because a large $\gamma_d$ heavily penalizes schedules with a high false positive rate. Thus, strategies that do not take false positives into account (e.g., \ba, \uten, and \uall) perform very poorly for high values of $\gamma_d$, whereas strategies that do take this cost into account (\dbr and \rbr) are much less affected.

\begin{tcolorbox}
\textbf{RQ2:} Varying $\gamma_d$ and $\gamma_a$ does not qualitatively affect our conclusion that \rbr is optimal compared to all other strategies for all considered settings of these parameters.
\end{tcolorbox}

\begin{table}
\centering
\caption{Randomized best response: schedules and probabilities ($\gamma_a = 1$, $\gamma_d = 2$). For each budget, we show the defender utility, chosen schedules, and respective probability for choosing each schedule.}
\begin{tabularx}{\columnwidth}{crXr}
\toprule
\textbf{Budget} & \textbf{Utility} & \textbf{Schedule} & \textbf{Probability} \\ \midrule
\multirow{6}{*}{1} & \multirow{6}{*}{-2.341} & Avira & 0.185 \\
& & ClamAV & 0.024 \\
& & ESET NOD32 & 0.029 \\
& & F-Secure & 0.019 \\
& & TrendMicro & 0.119 \\
& & ZoneAlarm & 0.625 \\ \midrule
\multirow{4}{*}{2} & \multirow{4}{*}{-2.270} & ClamAV, F-Prot & 0.237 \\
& & DrWeb, Kaspersky & 0.322 \\
& & Kaspersky, TACHYON & 0.119 \\
& & TheHacker, ZoneAlarm & 0.322 \\ \midrule
\multirow{4}{*}{3} & \multirow{4}{*}{-2.269} & Avast-Mobile, ClamAV, F-Prot & 0.237 \\
& & DrWeb, Kaspersky, TheHacker & 0.203 \\
& & DrWeb, Kaspersky, TotalDefense & 0.441 \\
& & Kaspersky, SUPERAntiSpyware, TheHacker & 0.119 \\ \midrule
\multirow{2}{*}{4} & \multirow{2}{*}{-2.269} & Avast-Mobile, ClamAV, F-Prot & 0.237 \\
& & DrWeb, Kaspersky, TheHacker, TotalDefense & 0.763 \\ \bottomrule
\end{tabularx}
\label{table:r_br_schedules}
\end{table}

\subsection{Resource Use} \label{sec:run_time}

\begin{table*}
\centering
\caption{The runtime of all the stages in our pipeline (in seconds). Load parameters is the time needed to load model parameters from VirusTotal and generate utility matrices for the defender and attacker. Generate MILP is the time needed to write the MILP to a file. Read MILP is the time to read the MILP into CPLEX. CPLEX Presolve and Probe are CPLEX optimization stages that reduce the size of the MILP before solving. Solve is the wall time to solve the MILP. }
\resizebox{\textwidth}{!}{
\begin{tabular}{lcrrrrrr}
\toprule
\textbf{Budget} & \textbf{Load Parameters}  & \textbf{Generate MILP} & \textbf{CPLEX Read MILP}   & \textbf{CPLEX Presolve} & \textbf{CPLEX Probe} & \textbf{CPLEX Solve} \\
\midrule
1      & 0.397 $\pm$ 0.021 &0.055$\pm$ 0.003    & 0.010 $\pm$ 0.000  & 0.035 $\pm$ 0.009 & 0.006 $\pm$ 0.009 & 0.207 $\pm$ 0.110  \\
    2      & 0.408 $\pm$ 0.019 & 1.759 $\pm$ 0.118   & 0.245 $\pm$ 0.016  &1.306 $\pm$ 0.290 & 0.093 $\pm$ 0.025 & 3.447 $\pm$ 1.349 \\
    3      & 0.916 $\pm$ 0.033 & 36.934 $\pm$ 4.855 & 5.562 $\pm$ 0.921 & 32.118 $\pm$ 8.573 & 0.217 $\pm$ 0.064 & 66.400 $\pm$ 32.553  \\ 
    4      & 14.313 $\pm$ 0.691 & 523.592 $\pm$ 91.450 & 84.752 $\pm$ 20.274 & 457.395 $\pm$ 135.952 & 1.249 $\pm$ 0.384 & 873.180 $\pm$ 391.370  \\ \bottomrule

\end{tabular}}
\label{table:run_time}
\end{table*}

%

To evaluate the performance of our proposed \rbr strategy, we have measured its runtime as well its memory consumption. Computing optimal strategies uses the following pipeline:

\begin{enumerate}\label{list:runtime_params}
    \item \textbf{Load Parameters:} Load model parameters from VirusTotal dataset and generate utility matrices $u_d$ and $u_a$.
    \item  \textbf{Generate MILP:} Generate and write the MILP to a file.
    \item   \textbf{CPLEX Read:} CPLEX reads the MILP.
    \item\textbf{CPLEX Presolve/CPLEX Probing:} CPLEX applies optimizations which reduce the size and complexity of the MILP.
    \item \textbf{CPLEX Solve:} CPLEX solves the reduced-size MILP and outputs a solution. 
\end{enumerate}

Averaged over 12 runs per budget, \tabref{table:run_time} gives the runtime results for each of the following stages in our pipeline. The number of schedules is exponential in the size of the budget; a budget of $b=1$ has 86 possible schedules, $b=2$ has 3,741, $b=3$ has 106,081 and $b=4$ has 2,229,636. As a result, the runtime of each stage in determining optimal randomization also increases exponentially as tool budget increases. Determining optimal randomization for $b\in \{1,2,3\}$ can be computed efficiently, in a matter of seconds or minutes, but $b=4$ requires roughly \textasciitilde 0.5 hours to compute. 

 We further show the runtimes for computing baseline strategies in \tabref{table:baseline_rt} Since the baseline strategies do not require an MILP to solve them, they are computed in fractions of a second, compared to solving for \rbr, which takes minutes. Therefore, across all budgets, calculating the baseline strategies is 179--889$\times$ faster than calculating the strategy in our proposed model.

While the baseline have a much faster runtime for computing optimal strategies, this does not make our model prohibitively expensive to implement. Once the optimal randomization \rbr is computed, sampling from the distribution at runtime (e.g., when a file must be scanned) can be done in constant time. If enough files are scanned, the amortized runtime will be comparable to the baselines.

\begin{table}
\centering
\caption{Runtime to solve games for baseline strategies, averaged over 12 runs. Baselines share parameter loading,  which has a runtime of  $0.3967 \pm 0.0205 $ for budget 1, $0.4079 \pm 0.0191$ for budget 2, $0.9158 \pm 0.0334$ for budget 3 and $14.3128 \pm 0.6912$ for budget 4. We detail the baseline strategy abbreviations in \secref{labeldescriptions}.}
\begin{tabular}{lcr}
\toprule
\textbf{Baseline Strategy} & \textbf{Budget}  & \textbf{Solve Time (s)}  \\
\midrule
\multirow{3}{*}{\ba} & 1 & 0.00024  $\pm$  0.00003 \\
& 2 & 0.00121 $\pm$ 0.00031 \\
& 3 & 0.03064 $\pm$  0.00588 \\
& 4 & 0.69511 $\pm$  0.09305 \\\midrule
\multirow{3}{*}{\uten} & 1 & 0.0003  $\pm$ 0.00002 \\
& 2 & 0.00093  $\pm$ 0.00002 \\
& 3 & 0.02670  $\pm$ 0.00034 \\
& 4 & 0.65883 $\pm$ 0.07263 \\\midrule
\multirow{3}{*}{\uall} & 1 & 0.00017 $\pm$ 0.00002 \\
& 2 & 0.00116  $\pm$ 0.00004 \\
& 3 & 0.02723 $\pm$ 0.00185 \\
& 4 & 0.62969 $\pm$0.13856 \\\midrule
\multirow{3}{*}{\eone} & 1& 0.00066 $\pm$ 0.00023 \\
& 2 & 0.00225  $\pm$ 0.00055 \\
& 3 & 0.04842  $\pm$ 0.00077 \\ 
& 4 & 1.12192  $\pm$ 0.13744 \\ \midrule
\multirow{3}{*}{\eten} & 1& 0.00037 $\pm$ 0.00002 \\
& 2 & 0.00165  $\pm$ 0.00007 \\
& 3 & 0.04780 $\pm$  0.00122 \\ 
& 4 & 1.12371 $\pm$  0.12602 \\ \midrule
\multirow{3}{*}{\dbr} & 1& 0.00012 $\pm$  0.00002 \\
& 2 & 0.00120  $\pm $ 0.00048 \\
& 3 & 0.03026  $\pm$  0.00408 \\
& 4 & 0.71663  $\pm$ 0.12118 \\\bottomrule
\end{tabular}
\label{table:baseline_rt}
\end{table}

To evaluate the memory cost of generating and solving these models, we measured the peak RAM usage for each budget size, for both the generation and solving of the MILP. For MILP generation, budgets 1, 2, 3, and 4 used an average of 147,244 $\pm$ 132 kb , 171,827 $\pm$ 254 kb, 1,455,271 $\pm$ 92 kb  and 28,664,755 $\pm$ 170 kb respectively. For MILP solving, budgets 1, 2, 3 and 4 used an average of 45,612  $\pm$ 136 kb, 572,383 $\pm$ 659 kb, 7,490,626 $\pm$ 50,617 kb and 95,986,773 $\pm$ 20,583 kb  respectively. We recorded these numbers using the \code{time} Linux command to record the peak RAM usage of each budget three times for both generating and solving each MILP. Due to the low variance for all of these values, we determined that four runs are sufficient for obtaining precise measurements.

Due to the number of schedules our MILP  compares, it faces a high memory cost when computing the optimal distribution. Fortunately, this high cost is only required once, as once an optimal distribution is calculated, sampling from it is trivial.


\section{Related Work}
\label{sec:related}
The work related to this paper is from three domains; security games, game theory for cybersecurity, and malware detection.

\subsection{Security Games}
Over the last decade, Stackelberg security games have been applied to a wide variety of security domains.  \citet{sinha2018stackelberg} provide an excellent overview of the field.  In this section, we highlight two particularly relevant works.

Our work is inspired by the work of \citet{jain}, which employs the use of Stackelberg security games to assign security resources in the LAX airport. Their work innovates by considering the different values for defense targets. While our work incorporates many ideas from their paper, we also present several improvements to their model. The first, and most important difference is that they apply their model for security in a traditional, physical setting, whereas our approach is for a more abstract defense. Attacks in the domain of malware detection are much lower risk for an attacker to make compared to terror attacks carried out in-person. Therefore, our domain, naturally, has a higher ratio of attackers to defenders. Additionally, while their model considers the several factors when judging the value of attacking certain targets, it does not consider the difficulty of carrying out an attack on the target, which our model incorporates.

To model the assignment of transit police patrol scheduling, \citet{yin} applied security problems on the domain of graph patrolling.
Like our own work, but unlike the work of \citet{jain} which considers terrorist attacks, \citeauthor{yin} model a domain in which attacks are relatively low-cost, and thus may frequently succeed.
Since attackers are commuters with routines, they plausibly will only change their decision of whether or not to buy a ticket, whereas attackers in our model may always change the type of their attack.  



\subsection{Game Theory for Cybersecurity}
Game theory and security games have historically been used to optimize cybersecurity defenses. In our discussion, we will focus more on the application of game theory to digital defenses than how to more effectively detect new malware.

\citet{10.1145/2808475.2808476} present a system for designing metrics for modeling non-static software defense. Our work expands on their model by adding costs for attackers and defenders, making the game non-zero sum. This change allows our model to avoid computing sub-optimal strategies as a result of ignoring these costs.

\citet{chung2016monitoring} criticize the use of game theoretic approaches to malware detection, including vulnerability to novel attacks, and a lack of a true measure of costs and rewards. The authors present a game theoretic approach for automated responses to network breaches, allowing systems to defend themselves without the need for human administrator action. Their approach specializes in domains where knowledge of attacks and their payoffs is limited, which is not the case in our work.

\citet{wang2014adhoc} present a many-player game theoretic approach to security in mobile ad-hoc networks (MANETs). Their model allows normal users in MANETs to make distributed security defense decisions, allowing the MANET to be more resilient to being compromised, while also minimizing the resources required to do so. Their work uses system resource constraints as a budget, which is similar to our use of budgets as a constraining factor.





Game theory has also been employed in ``honeypotting,'' where vulnerable ``bait'' systems are distributed to hide real ones, to waste attacker time, warn against incoming attacks, and understand new vulnerability exploits. \citet{Radek2012honey} uses game theory to optimally decide how vulnerable to make the bait systems, as well as how many to use, to maximize their effectiveness. Their later work \cite{kiekintveld2015honey} considers attackers who use probes to detect if a system is real, and uses attack graphs to better model an attacker's knowledge of a system. Attack probes explain why attackers can be best responding (by having perfect information) in our model, as an attacker is able to repeatedly probe our defenses to see which attacks are more likely to succeed.

\subsection{Malware Detection}
In the vein of malware detection, \citet{Measuring} provide insights about the VirusTotal~\cite{VT} dataset. The authors suggest that tool dependencies have a high bearing on the results that tools give, and that the top tools may be clustered into closely related sets.

\citet{Zolotukhin2013malware} argue that signature-based malware detection and classification is flawed, because it requires systems to be compromised before new signatures can be detected. Their alternative to this approach is to first analyze the bytecode of known malicious samples and look for patterns that are later used to detect new malware exhibiting similar behavior but possibly carrying a different signature. Their work studies how to find novel forms of malware, whereas our work focuses on effectively using existing tools to detect malware.

Much other research in malware detection has explored using machine learning techniques. These solutions seek optimal use of limited defensive resources, and are thus a complementary approach to ours. For example, to prioritize attacks that pose larger threats to an enterprise, \citet{MADE} use supervised learning to analyze security logs. \citet{Sharif2018PredictingIE} use neural networks to predict the likelihood of a user encountering malicious web domains and take preventative measures. \citet{Stefanova2018OffPolicyQT} use reinforcement learning to detect and respond optimally to network attacks. 
\citet{xiao2017phones} studies the problem of balancing the high computing cost of scanning mobile devices for malware, with the low computing and battery power associated with the devices. They use cloud computing to analyze a documentation of the system state. They present two malware detection strategies; Dyna-Q, which offers faster scans, and PDS, which offers higher malware detection accuracy.

\section{Conclusion}
\label{sec:conclusion}

There are a plethora of malware detection tools available, each with their own strengths and weaknesses. To improve coverage, an organization may therefore wish to use a suite of tools to defend against potential malware attacks, which immediately raises the question of which subset to run.  A subset that is large enough to detect every attack would potentially require more computing resources and incur more financial costs than feasible.  However, attackers are strategic. If the organization deterministically chooses any subset of tools that leave some vulnerabilities open, then those vulnerabilities will be preferentially attacked.  The solution is to randomly select a subset of tools to run.  In this work, we have presented a novel approach for computing an optimal randomization over feasible sets of malware detection tools that incorporates the costs for both the attacker and the defender.  Our empirical evaluation shows that our best-responding randomization strategy outperforms deterministic strategies, as well as naive randomization strategies.

The defender's costs depend on some factors that are objective (e.g., false positive rates for different tools on different attacks) and others that are subjective (e.g., relative costs of different vulnerabilities' being exploited).  For this work, we estimated objective factors from past studies \citep{Measuring} and the VirusTotal dataset, and chose values for subjective factors about the severity of attacks using standard ratings in the National Vulnerability Dataset.  However, an advantage of our approach is that the utility model can be customized to reflect the individual costs, preferences, and available tools for each specific domain to which it is applied.

It is less straightforward to estimate the attacker's utility than the defender's.
A defender who wishes to take a conservative approach can assign a utility to the attacker that is equal to the severity of each attack for the defender.  If partial information is known about the attacker's utility, then it can be incorporated into the model as well.  Our approach does not require full knowledge of the attacker's utility; rather, it takes advantage of whatever knowledge might be available to the defender.

The process of finding and exploiting vulnerabilities is fundamentally strategic.
To the best of our knowledge, our work is the first to explicitly account for strategic attacker behavior while optimizing the tradeoff between costs and benefits in the malware detection domain.

\bibliographystyle{ACM-Reference-Format}
\bibliography{gt-security.bib}


\end{document}